\documentclass[12pt]{article}
\title{Size reduction and partial decoupling of systems of
equations\footnote{This work was partially supported by the Natural
Sciences and Engineering Research Council of Canada under CRD Grant 
\# 661-090/93 and partially supported by the Engineering and Physical
Sciences Research Council England under Grant \#  GR/L99036}}
\author{Thomas Wolf\\ Department of Mathematics, Brock University, \\
        500 Glenridge Avenue, St.\ Catharines, 
        Ontario, Canada L2S 3A1 \\  email: twolf@brocku.ca}

\begin{document}

\maketitle
\begin{abstract}
A method is presented that reduces the number of terms of systems of
linear equations (algebraic, ordinary and partial differential
equations). As a byproduct these systems have a tendency to become
partially decoupled and are more likely to be factorizable or
integrable. A variation of this method is applicable to non-linear
systems. Modifications to improve efficiency are given and examples
are shown. This procedure can be used in connection with the
computation of the radical of a differential ideal (differential
Gr\"{o}bner basis).
\end{abstract}

\section{Motivation}
Algorithms for applying integrability conditions to a system of
differential equations in a systematic way in order to 
generate simplified differential equations are implemented in a
number of programs (\cite{BLOP1, BLOP2, Hub, Hub2, Mans, Reid2, Reid3} 
and more in \cite{WHER}). Such calculations result in the
radical or a (pseudo) differential 
Gr\"{o}bner Basis of the differential ideal generated by the 
original system. A common problem of these algorithms, and
consequently their implementations, is an explosive expression
swell. Optimizations like Buchberger's 
$2^{\mbox{\footnotesize nd}}$ criterion 
(section 5.5 in \cite{BW}) and their
analogue for differential equations aim to reduce the number of steps to
reach a characteristic set
(one step = computation of an S-polynomial
for algebraic systems or a cross-differentiation of two
differential equations for differential systems). These optimizations
do not cover other `obvious' simplifications.
For a very simple example,
consider $Df$ to be a leading derivative of a function $f$ 
in two equations $0 = Df + A$, $0 = Df + 2A$
with $A$ a sum of a large number of terms. A simplification step
in the standard procedure would aim at eliminating $Df$ and get as a
consequence the system
$0 = Df + A, 0 = A$ where the big expression $A$ occurs twice. 
An alternative to be described in this paper would be to
try to shorten equations and therefore to get at first 
the system $0 = Df + A$, $0 = Df$ and in a second length reduction step the
system $0 = A$, $0 = Df$. 

For a slightly more realistic example, consider two equations
$0=A+C$ and $0=B-C$, where $A,B,C$ are differential expressions, $A,B$
having only few terms and $C$ involving many terms. If both equations
are long and if they involve high derivatives then they would have a
low priority to be used in standard algorithms. Typically, each of
them would be paired individually with short low order equations for
their reduction or the generation of integrability conditions. A
simplification of both equations to the system $0=A+C$ and $0=A+B$
would usually not be found. Instead the expression $C$ would grow in
{\it both} equations when any substitutions of functions are made that
occur in $C$.

Both examples are clear cut situations where big expressions in
different equations can cancel each other. Although in practical
applications to be described in a later section each length reduction
step saves only a small number of terms, these reduction steps can
often be repeated many times. 

The need to reduce the length of equations comes from the danger of
dealing with long equations in elimination algorithms. For elimination
algorithms to be finite, i.e.\ to involve only a finite number of
steps, they have to eliminate the leading derivative 
of equations {\it first} which involves differentiations and
multiplications. Both tend to increase the length of long equations
even further. However, elimination algorithms would remain finite if
terms other than the leading derivative would be eliminated a {\it finite}
number of times. Flexibility of this kind could be used to prevent
excessive expression swell. In this way not only memory is saved.
Long expressions also require an increased time to be computed which
slows down any future computations in which long expressions are
involved. 

The procedure to be described in section \ref{proc} is a first step in
the direction of an `intelligent' and more efficient computation.
It aims at finding equations in the algebraic ideal of
the given system with fewer terms. The basic idea is rather
straightforward. Any pair of two equations of a given system of
equations is checked whether there is a linear combination 
(with non-vanishing coefficients) of these
two equations that is shorter than the longer of the two. If that is
the case then the longer one is replaced by the shorter new
equation. To find a length reducing linear combination of a pair of
two equations, each term of one equation is divided by each term of
the other equation, the quotient is simplified, i.e.\ common factors
of numerator and denominator are dropped and a counter of the number
of occurrences of this quotient is incremented.
The quotient occurring most often is
picked, its numerator and its denominator are the multipliers of both
equations which are then subtracted from each other. By choosing the
quotient that occurs most frequently a maximum number of terms cancel
and the result is as short as possible. 

The main content of the first part of the paper is to introduce data
structures ($L, c_i$ below) which allow an efficient implementation
and to describe some optimizations that speed up the method
and that allow one to consider non-linear equations for length reduction. 
Beneficial side-effects are discussed such as the increased chance to
find ordinary differential equations (ODEs) or to find 
exact differential equations in a length reduced system of
partial differential equations (PDEs). 

In the second half, in section \ref{exmpl} the method is applied to
determining equations of Killing vectors and Killing tensors in
General Relativity. One example is explained in more detail in the
appendix where the beneficial side-effects of length reduction become
important. 


The length reduction module is incorporated in the package {\sc Crack}
for solving over determined PDE-systems. To show that the usefulness
of the length reduction module is not just based on special features
of {\sc Crack} but is of a more universal nature a test is described
at the end of section \ref{exmpl}. In this test other well known programs are
used to solve a system of PDEs before and after it has been length
reduced. The suitability of length reduction as a pre-processing step is
demonstrated.  

\section{The term reduction method}  \label{proc}
{\bf \ Introduction}\vspace{3pt}\\ 
The procedure to be described takes as input two
expressions $E_1, E_2$. Both represent equations $0=E_1=E_2$
and are regarded as sums with $n_1$ and $n_2$ terms ($n_1,n_2\geq 1$).
The aim is to multiply each with a different single term and to add the
expressions such
that the result has fewer terms than max($n_1,n_2$) and can be substituted
for the longer of $E_1, E_2$. For a given system of equations this
process is repeated with all possible pairs of equations until no pair
can produce an equation that is shorter than the longer of both. 
The restrictions to multiply only with monomials and to combine only
two equations at a time are non-trivial constraints
(see the discussion in section \ref{PosEx}). 

{\it Example:}
The method will be illustrated with the following two expressions
\begin{eqnarray}
E_1 & = & 2xf + 6yf + 4xg + 5x = \sum_{i=1}^4 E_{1i} \label{ex1} \\
E_2 & = & 3yf - 3xf + 6yg - 7y = \sum_{i=1}^4 E_{2i} \label{ex2}
\end{eqnarray}
$f$ and $g$ are the unknowns that are to be computed from $0=E_1,
0=E_2$. To explain the method it does not matter if $f$
and $g$ would be replaced by derivatives of unknown functions or by
products of powers of different derivatives as long as $f \neq g$.
$x$ and $y$ are independent variables or parameters such that $0=E_1,
0=E_2$ are to be satisfied for any value of $x$ and $y$.

{\vspace{6pt} \bf The treatment of two equations}\vspace{3pt}\\ 
Given are two expressions $E_1, E_2$
with $n_1$ and $n_2$ terms, $n_1 \geq n_2$. If each expression is
multiplied with a single term (monomial) and both expressions are added, then
their sum $E_3$ can have between $n_1-n_2$ and $n_1+n_2$ terms
depending on how many terms cancel each other. A way to find the
optimal cancellation, i.e.\ optimal multipliers is to
divide each term of expression $E_1$ by each term of 
$E_2$ and to collect the simplified quotients (common factors of
numerator and denominator dropped) together with the multiplicity
they occur.

{\it Example:} $E_1, E_2$ given in (\ref{ex1}),(\ref{ex2})
have 4 terms each ($n_1=n_2=4$) and the quotients are
\[ \left(E_{1i}/E_{2j}\right) = \left(
\begin{array}{cccc}
{2x \over 3y} & -{2 \over 3} & {xf \over 3yg} & -{2xf \over 7y} \\ & & & \\
2 & -{2y \over x} & {f \over g} & -{6f \over 7} \\ & & & \\
{4xg \over 3yf} & -{4g \over 3f} & {2x \over 3y} & -{4xg \over 7y}  \\ & & & \\
{5x \over 3yf} & -{5 \over 3f} & {5x \over 6yg} & -{5x \over 7y}  
\end{array}
\right). \]
A new equation $E_3$ would be generated by picking a quotient, say
$E_{12}/E_{21}={4xg \over 3yf}$ and using its numerator and
denominator to compute $E_3=3yf\cdot E_1 - 4xg\cdot E_2$.
Because this quotient involves $f$ and $g$, the new equation $0=E_3$ is
non-linear in $f, g$. As a consequence after replacing $E_1$ by $E_3$
the new system $0=E_3=E_2$ may not be equivalent to the old system
$0=E_1=E_2$, i.e.\ $E_1$ could not automatically be replaced by
$E_3$. The algorithm will therefore not consider quotients that
involve any unknowns, here $f$ and $g$. An effective method to avoid
the computation of such quotients will be explained further below.

The quotient occuring most often is ${2x \over 3y}$ which is appearing
twice and is free of $f$ and $g$. Each appearance of a quotient
means that two terms cancel. We therefore would expect that in
the expression  
$E_3:=3y\cdot E_1 - 2x\cdot E_2$ two times two terms cancel and
therefore $E_3$ has $4 + 4 - 2\cdot 2 = 4$ terms. But 
$E_3=6fx^2 + 18fy^2 + 29xy$ has only 3 terms. When computing
$E_3:=3y\cdot E_1 - 2x\cdot E_2$ the two terms $3y\cdot E_{14}-2x\cdot
E_{24}= 3y\cdot 5x-2x\cdot (-7y)=15xy+14xy$ add up to only one term $29xy$.
We could have forecast the saving of one additional term by realizing that
the quotient $E_{14}/E_{24}= -{5x \over 7y}$ differs from
$E_{11}/E_{21}= {2x \over 3y}$ only by a numerical factor. 

The example implies that we should record all quotients with the
multiplicity they occur and group them into classes $c_i$
where all quotients in a class differ by only a numerical factor.
Finally, for each class $c_i$ the sum $M_i$ of all the multiplicities of all
quotients in the class is recorded too. All classes $c_i$ together
with $M_i$ are listed in a list $L$:
\begin{eqnarray*}
L   & = & ((c_1,M_1), (c_2,M_2),\ldots,(c_r,M_r)) \\
c_i & = & ((q_{i1},m_{i1}), (q_{i2},m_{i2}),\ldots,(q_{is_i},m_{is_i})).
\end{eqnarray*}
\begin{itemize}
\item[$q_{ij}$] are the different quotients such that two quotients $q_{ij},
q_{ik}$ in the same class $c_i$ differ only by a numerical factor.
\item[$s_i$] is the number of different quotients in the class $c_i$.
\item[$m_{ij}$] is the number of how often $q_{ij}$ occurs. 
\item[$M_i$] are defined as $M_i = \sum_{j=1}^{s_i} m_{ij}$.
\item[$L$] is the complete (unsorted) list of all classes of quotients.
\end{itemize}
Disregarding quotients involving $f$ or $g$ in the above example we have
\begin{tabbing}
\ \ \ \ \ \ \ \ \ \ \ \ \ \ \ \ \ \ \ \ \ \ \ \ \ \ \ \
$L =$ \= (\=$((({2x \over 3y},2), (-{5x \over 7y},1)), 3),$ \\
      \>  \> \\
      \>  \>$(((-{2 \over 3},1), (2,1)), 2),$ \\
      \>  \> \\
      \>  \>$(((-{2y \over x},1)), 1)$ \\
      \>  \> \\
      \> )\>,
\end{tabbing}
with, for example, 
$q_{11}={2x \over 3y}$
turning up twice, once as $E_{11}/E_{21}$ and once as $E_{13}/E_{23}$,
therefore $m_{11}=2$, and further
$c_1 = (({2x \over 3y},2), (-{5x \over 7y},1)), \; 
 M_1 = m_{11} + m_{12} = 2 + 1 = 3$. 

If a quotient $q_{ij}$ is used to combine $E_1,E_2$ to
\[ E_3 = \mbox{denominator}(q_{ij})\times E_1 - 
         \mbox{numerator}(q_{ij})\times E_2\]
then the number $n_3$ of terms of $E_3$ is 
\begin{eqnarray}
   n_3 & = & n_1+n_2-2\times m_{ij} \; 
             \;\;\;\;\;(\mbox{due to $m_{ij}$ complete cancellations of 2 terms
             } )  \nonumber \\
       &   & \;\;\;\;\;\;\;\;\;\;\;\;
             -\sum_{k=1,k\neq j}^{s_i} m_{ik} 
              \;\;(\mbox{due to savings of one term each time}) \label{n3} \\
       & = & n_1+n_2-m_{ij}-M_i.                     \nonumber
\end{eqnarray} 
The $\sum m_{ik}$ in eqn.\ (\ref{n3}) comes
from simplifications like $15xy+14xy=29xy$ which each save one
term. In order to be successful and to replace $E_1$ by $E_3$ 
we need to find a quotient $q_{ij}$ such that $E_3$ has fewer terms
than $E_1$, i.e.\ $n_3=n_1+n_2-m_{ij}-M_i < n_1$, hence $m_{ij}+M_i>n_2$.
\vspace{6pt}

{\bf A pre-processing step}\vspace{3pt}\\
As argued above only quotients $q_{ij}$ should be considered which do
not involve any unknowns, i.e.\ functions or constants that are to be
computed from $E_i=0$. An effective method to even avoid the
computation of those quotients requires to re-write expressions $E_1,
E_2$ in the following way. This initial re-writing step also clarifies how
the method works for non-linear expressions $E_1, E_2$.

One always can regard non-linear expressions 
$E_1,E_2$ as linear homogeneous expressions in some
newly defined variables $v_a$ which are linear or non-linear
constructs of the dependent variables. For example, for independent
variables $x,y$, dependent variables $f=$ const, $g=g(x,y)$ and
\begin{eqnarray}
 E_1&=&3x+3\cos(x)fg_x-12xyg+6xg-yg+\sin(g_y),    \label{li1} \\
 E_2&=&1+4fg_x-4yg+2g                             \label{li2}
\end{eqnarray}
the related system that is homogeneous and linear in $v_i$ would be
\begin{eqnarray}
 E_1&=&3xv_0+3\cos(x)v_1-12xyv_2+6xv_2-yv_2+v_3,  \label{ls1} \\
 E_2&=&v_0+4v_1-4yv_2+2v_2                        \label{ls2}
\end{eqnarray}
with $v_0=1,\;\;v_1=fg_x,\;\;v_2=g,$ and $v_3=\sin(g_y)$.
After this re-writing the method will
investigate the system (\ref{ls1}),(\ref{ls2}) and avoid quotients
$q_{ij}$ that involve any $v_l$. 
\vspace{6pt}

{\bf Methods to increase speed}\vspace{3pt}\\ 
The restriction of not multiplying with factors involving dependent
variables enables the following major speed up.
Instead of investigating the system (\ref{li1}),(\ref{li2}) and
computing $6\times 4 = 24$ quotients we investigate
the system (\ref{ls1}),(\ref{ls2}) and compute only quotients
between the terms of the coefficients of the same $v_i$ in $E_1$ and
$E_2$. This reduces the number of quotients to
$1\times 1$ (for $v_0$) $ + 1\times 1$ (for $v_1$)
$ + 3\times 2$ (for $v_2$) $ + 3\times 0$ (for $v_3$) $ = 8$ quotients.
For large expressions, or more exactly for a high number
of different $v_j$  the speed up is naturally much higher.
If we have $r+1$ new dependent variables $v_0,\ldots,v_r$ and 
if we denote the number of terms involving $v_j$ in $E_i$ as $n_{ij},
\;\; 0\leq j \leq r$ then instead of computing 
$\left(\sum_{j=0}^r n_{1j} \right)\times \left(\sum_{k=0}^r n_{2k} \right)$
quotients the more efficient method only computes
$\sum_{j=0}^r \left(n_{1j}\times n_{2j}\right)$ quotients.

Another way of increasing efficiency is based on knowing the
$n_{ij}$ beforehand. For the terms involving a specific $v_j$, an
upper bound on the maximal number of cancellations is
min($n_{1j},n_{2j}$), saving twice as many terms. This value summed
over $j=0\ldots r$ (for each $v_j$) gives an upper bound on how many
terms can be saved due to cancellations.

This test can be performed without computing any quotients:\\
{\em If the new dependent variables are
$v_0,\ldots,v_r$ and if we have at the beginning 
\[ \sum_{j=0}^r 2\,\mbox{min}(n_{1j},n_{2j}) \leq 
   n_2 \;\;\left(\;= \sum_{j=0}^r n_{2j}\right) \]
($n_2+1$ is the minimum number of terms to be saved to reach a length
reduction) then no length reduction is possible.}

This test of a necessary criterion
is not only possible at the beginning
but also during the computation of
quotients. We assume that at first all quotients related to $v_0$ are
computed, then those related to $v_1$, and so on.
When the calculation has reached $v_j$ and 
the first $w$ terms in $E_1$ that involve $v_j$ have been processed,
i.e.\ all quotients between each of them and all terms in $E_2$ with $v_j$
have already been computed
then the following holds. At most min($n_{1j}-w,n_{2j}$) more
cancellations related to $v_j$ are possible and 
an upper bound $B_j(w)$ 
of the total number of any cancellations still to be found is given as 
\[ B_j(w) := \mbox{min}(n_{1j}-w,n_{2j}) + 
                      \sum_{i=j+1}^r \mbox{min}(n_{1i},n_{2i}). \]
At this moment any quotient $q_{kl}$ that has a chance to
provide a length reduction must satisfy 
\[M_k+m_{kl} + 2B_j(w) > n_2. \]
All $q_{kl}$ which do not satisfy this condition can be dropped from
the list $L$. For the same reason no new quotients should be added to
$L$ as soon as
$2B_j(w) \leq n_2$. If the list $L$
becomes empty at any time during
the computation of quotients, then the search
can stop. In that case no length reduction is possible. By dropping
quotients from the list $L$, the updating of $m_{ij}$ and $M_i$ speeds
up.\vspace{6pt}

{\bf A speed up not recommended} \vspace{3pt}\\ 
When deleting quotients from $L$ that have no chance to give a length
reduction then it is little extra effort to check whether any $q_{kl}$
of the remaining quotients in $L$ 
already satisfies $M_k+m_{kl} > n_2$, and therefore is guaranteed
to provide a length reduction. As soon as such a quotient $q_{kl}$ is found
the execution could stop. 
In practical tests it appeared that the negative effects of an early stop
dominate. The first length reducing quotient $q_{kl}$ that is found does not
have to be the one giving the highest length reduction possible.
After using a sub-optimal $q_{kl}$ to compute $E_3$ and to substitute $E_1$
no further length reductions may be possible, or even if further
length reductions were possible, 
equations tend to have at least intermediately more terms compared with
determining always the optimal quotient that gives the highest 
length reduction. After a suboptimal length reduction, subsequent 
pairings with other equations would be slower which would result in an
overall slow down. It therefore is
recommended to complete the computation of all relevant quotients and
not to stop early when the first length reducing $q_{kl}$ is found.
\vspace{6pt}

{\bf Some time tests} \vspace{3pt}\\ 
The following tests can only provide some idea of the running times of
the length reduction method. 
They are measured in a 8 MB {\sc Reduce 3.6} session running under {\sc Linux}
on a 133 MHz Pentium PC (Dec.\ 1998).
Equations which have been paired had been
generated with the {\sc Reduce} command {\tt RANDPOLY} which allows 
one to specify the number of terms to be generated.
In {\tt RANDPOLY} the randomly generated coefficients may become zero which 
in that case results in a polynomial with fewer
terms than specified. Therefore
the number of terms was chosen somewhat larger to be able
to drop the surplus terms and get the required size of the polynomial
and perform the following statistics.

\begin{figure}[htbp]
 \input{tst} 
\newline  \indent Figure 1. Running times for random polynomials of
different sizes.
\end{figure}

Performing a length reduction investigation involves no other risk or
cost than the computer time that may be lost if no length reduction
was possible. Therefore in figure 1 and table 1 a statistics of
investigations of each time two equations is shown where no length
reduction could be found, i.e.\ the worst possible result. 
This will be referred to as unsuccessful
pairings in contrast to successful pairings where a length reduction
was possible.
Both equations are random polynomials of degree up to 5 and with up to 8
variables. Times are obtained by averaging 20 runs. The individual
times in these runs differ typically by up to 30\%. The results
confirm an overall  
dependence \ \ \ time $\propto$ (terms of eqn.\ 1)$\times$(terms of eqn.\ 2).
\small
\begin{center}
\begin{tabular}{|l|c|c|c|c|c|c|c|c|c|c|} \hline
\begin{tabular}{l}
no of terms of 1$^{\mbox{\footnotesize st }}$ eqn. 
\end{tabular} 
 & 100 & 200 & 300  & 400 & 500 & 600  &  700 &  800 &  900 & 1000 \\ \hline
\begin{tabular}{l}
time in sec (no success) \\
if 2$^{\mbox{\footnotesize nd }}$ eqn.\ has 10 terms 
\end{tabular} 
 & 0.07 & 0.23 & 0.45 & 0.70 & 0.72 & 1.01 & 1.36 & 1.8 & 2.2 & 2.7 \\ \hline
\begin{tabular}{l}
time in sec (no success) \\
if 2$^{\mbox{\footnotesize nd }}$ eqn.\ has 1000 terms 
\end{tabular} 
 & 4.18 & 6.23 & 9.35 & 12.6 & 15.4 & 20.3 & 24.4 & 30.2 & 36.4 & 43.8 
\end{tabular} 
\end{center}
\normalsize

Table 1. Timings of unsuccessful length reduction attempts of one
equation with varying length and a second equation that has either 10
or 1000 terms.\vspace{12pt} 

Test results shown in table 2 are based on pairing equations which are
polynomials of 7$^{\mbox{\footnotesize th }}$ degree with 
each 300 terms but with a varying number of variables.
Times are averaged again over 20 runs.

The effect of efficiency improvements as described in the above
sub-section to detect the non-existence of length reductions early can
be seen clearly from the second row in table 2. As more independent
variables occur the number of different quotients $q_{kl}$ increases
and the average frequency for each quotient to appear becomes
smaller. This in turn rules out many quotients early in the
computation and it becomes clear earlier that no quotient will result
in a length reduction if that is the case. 

Usually length reductions do not happen with random polynomials of that
size. In order to measure computing times for pairings
when length reductions were possible, 
pairs of polynomials had been constructed in the following way:
a multiple of one random polynomial $P_1$ of 300 terms is added to another
random polynomial $P_2$ and terms in excess of 300 terms are dropped
to obtain a polynomial $P_3$ with 300 terms. Length reductions between
polynomials $P_1$ and $P_3$ are investigated which produced the
third row in table 2. Two trends seem to
be present, one lowering the time with an increase of the number of
variables (mainly effective between 4 and 5 variables) due to a
decrease of potentially successful quotients and another trend slightly
increasing the time with the number of variables.
\begin{center}
\begin{tabular}{|l|c|c|c|c|c|} \hline
no of variables & 4 & 5 & 6 & 7 & 8  \\ \hline 
time in msec (unsuccessful)   & 4720 & 2461 & 1673 & 1656 & 1640 \\ \hline 
time in msec (successful)     & 5706 & 4968 & 5137 & 5388 & 5728 \\ \hline 
\end{tabular} 
\end{center}

Table 2: A comparison between average running times of unsuccessful and
successful pairings of two equations.\vspace{6pt} 

To summarize this sub-section,
the main feature found in this study is that computing times are lower
in unsuccessful attempts to shorten equations and only when a
length reduction becomes possible, i.e.\ when computing times are of less
importance, then they are increased. The speed up measures become
increasingly efficient if more unknowns (like $f$ and $g$ in the first
example) are present which is the case for partial differential
equations with many different partial derivatives acting as different
unknowns $v_j$ during length reduction. \vspace{6pt}

{\bf The order of pairings of equations}\vspace{3pt}\\ 
If more than two equations are given then the question arises in which
order they should be paired to search for length reductions.
Given that we combine only two equations at a time and
multiplying them only with a monomial, we can not expect results that
are invariant against combining equations in a different order.
The following criteria serve only as a suggestion but they
proved to be useful in applications. According to them
pairs of equations are picked with the following priorities:
\begin{itemize}
\item There should be as {\it few} as possible
dependent variables $v_i$ in the shorter equation
which do not occur in the longer equation.
\item The shorter of both equations should be as short as possible.
\item The longer of both equations should be as short as possible.
\end{itemize}
The first two rules maximizes the chance to find a reduction of terms.
The third rule reduces computation times. The second rule has a higher
priority than the third rule because the shorter the equations are,
the more useful they are potentially in reducing the length of other
equations. 

In the following table the above priority list
is compared with the same list,
only modified by exchanging in the first rule `as few as possible'
with `as many as possible'. The equations are a set of first order
partial differential equations (PDEs) resulting from investigating 
in General Relativity the Kimura metric \cite{Kim} with respect to Killing
tensors (see section \ref{exmpl}). Because sin and cos occur in these
differential equations, both length reductions are performed once
with the simplification rule $\cos(x)^2 \Rightarrow 1-\sin(x)^2$ and
once with the simplification rule $\sin(x)^2 \Rightarrow 1-\cos(x)^2$.
It becomes apparent that these simplifications 
are not equivalent in their effect.

\begin{center}
\begin{tabular}{|c||c|c||c|c||c|c|} \hline
simplification &\multicolumn{2}{c||}{original system}
               &\multicolumn{2}{c||}{length red.\ system}
               &\multicolumn{2}{c|}{length red.\ system}\\ \cline{2-7}
   used        &\multicolumn{2}{c||}{               }
               &\multicolumn{2}{c||}{using the rule}
               &\multicolumn{2}{c|}{using the rule}\\ \cline{2-7}
               &\multicolumn{2}{c||}{               }
               &\multicolumn{2}{c||}{`... as many as ...'}
               &\multicolumn{2}{c|}{`... as few as ...'}\\ \cline{2-7}
               & no of eqn & terms & no of eqn & terms & no of eqn & terms \\ 
\hline \hline
$\sin(x)^2 \Rightarrow 1-\cos(x)^2$ & 48 & 607 & 25 & 117 & 21 & 81 \\ \hline
$\cos(x)^2 \Rightarrow 1-\sin(x)^2$ & 48 & 464 & 25 & 108 & 21 & 74 \\ \hline
\end{tabular} \vspace{6pt}\\
\end{center}

Table 3: A comparison of different simplification rules and different
rules for the pairing of equations. \vspace{6pt}\\
As it was to be expected, the `... as few as ...' rule performed
better (i.e.\ resulted in shorter length reduced systems)
than the `... as many as ...' rule. What also becomes apparent
is that choosing accidentally the less effective simplification rule 
$\sin(x)^2 \Rightarrow 1-\cos(x)^2$ is less critical when length
reduction is performed than without length 
reduction: 
\begin{eqnarray*}
{\mbox{no of terms of original system using cos-rule} \over 
 \mbox{no of terms of original system using sin-rule}       }
      \; = \; {607 \over 464} & > & \\
{\mbox{no of terms of reduced system using cos-rule} \over 
 \mbox{no of terms of reduced system using sin-rule}       }
\; = \; {81 \over 74}. & &
\end{eqnarray*} 

{\bf Beneficial side effects}\vspace{3pt}\\ 
The reduction of length is not only useful for saving memory, and as
shorter expressions are quicker to process later on, also for saving time.
In this section we want to explain further benefits.
\begin{enumerate}
\item \label{B1}
  Given a set of PDEs, the length reduced system is more likely to
  contain ordinary differential equations (ODEs) or to contain
  integrable exact PDEs. 
\item \label{B2}
  Length reduced polynomially non-linear equations are much more
  likely to be algebraically factorizable.
\item \label{B3}
  Length reduction of a system of equations has in general the
  side effect of partially decoupling the system. 
\end{enumerate}
If the number of terms of an equation is
lowered to, say, $n$ and if the equation is linear then $n$ is
necessarily an upper bound for the number of different functions and
different derivatives that occur. The length reduction method as described
above is indiscriminate to different functions or different 
derivatives. On average therefore a consequence of
a reduction of the number of terms will be a reduction of the
number of different functions and the number of different derivatives
that occur (see the comparison between tables 6 and 7 below) which in
turn provides the above side effects. \vspace{6pt}

{\it About \ref{B1}:}\\
If the number of terms in a length reduced equation 
got very small (say less than 4) then the
chance increases that they involve only one differentiation variable,
or that all derivatives can be looked at as derivatives with respect
to only one variable of a common partial derivative, like
$\partial_y f, \partial_{xy} f, \partial_{xxy} f$ are all
$x$-derivatives of $\partial_y f$. In these cases the equation has been
reduced to an ODE. (A proper ODE can of course only appear if the differential
ideal of the original system does contain ODEs, but that is guaranteed
for all the typical sources of over determined PDE-systems, like the
computation of infinitesimal symmetries and of conservation laws.)

Similarly the chance increases to obtain an exact differential
equation. In order for a differential expression $P(f^i)$ that 
involves functions $f^i$ and that satisfies $0=P$ to be a total $x$
derivative of some expression $I(f^i)$, the identity $P=dI/dx$ has to
be satisfied identically in all functions $f^i$ and in all their
derivatives. The fewer different functions $f^i$ and the fewer
different derivatives with respect to variables other than $x$ occur,
the less restrictive is this assumption of exactness and the more
likely it will be satisfied.

The conclusion is not that length reduction is the best way to proceed in 
order to integrate. We only say that the chance increases to find an
integrable PDE in a length reduced system. In the appendix this statement 
is illustrated by a sequence of 10 integrations which a system of 
Killing tensor equations admits after being reduced in length.

To explain the usefulness of integrability let us consider an extreme 
hypothetical example.
Gr\"{o}bner Basis techniques aim at equations with a low differential 
order. This is a good strategy but not the only way to go. For example,
knowing that $0=\partial^{10}f/\partial x^{10}$ is included in the
differential ideal would be very useful as well. After integration,
substitution of $f$ and direct separation with respect to $x$
(sometimes called splitting or fragmentation) a highly over determined
system for the 10 functions of integration results. 
Although an expensive Gr\"{o}bner basis computation might have
provided an ODE of lower order than 10, this information is
usually gained faster by integrating 
$ 0 = \partial^{10}f/\partial x^{10} $ and
solving the over determined system that resulted from direct separation
for the 10 functions of integration. The key to this speed up would be to
sacrifice the minimal differential order for a reduction in the number
of terms. \vspace{6pt}

{\it About \ref{B2}:}\\
In a small experiment we want to show that the chance for an algebraic
factorization increases with a reduction of the number of terms. 
Bi-linear polynomials with up to 5 variables and
coefficients in the interval $-9,\ldots,-1,1\ldots 9$ have been
randomly generated for each length from 2 terms up to 6 terms. 
The percentage of factorizable polynomials is given in table 4.
The chance to have a non-trivial factorization
(non-numeric factors) decreases surprisingly quick with an
increasing number of terms. Although this test is not proving anything
it may serve as an illustration. 
\begin{center}
\begin{tabular}{|l|c|c|c|c|c|} \hline
no of terms & 2 & 3 & 4 & 5 & 6    \\ \hline
factorizable equations in \% & 55.4 & 13.6 & 2.3 & 0.22 
       & $1.3\times 10^{-3}$  \\ \hline
\end{tabular} 
\end{center}

Table 4: The chances for random quadratic homogeneous polynomials in up to 5
variables to be factorizable in dependence on the number of terms.\vspace{6pt}

{\it About \ref{B3}:}\\
The fewer different functions occur in each equation (on
average) the more the system is (at least partially) decoupled and the
fewer steps are needed in a subsequent computation to get a differential
Gr\"{o}bner Basis or characteristic system. This means that an
elimination algorithm performed afterwards has less work to do, i.e.\ needs
fewer steps. A sparsely occupied system also opens the possibility
to choose an appropriate total ordering on which an elimination
algorithm is based. For example, if a function turns up in only few
equations and with very few different derivatives then this function
should get a high priority, i.e.\ the lexicographical ordering of
functions which plays a role in any total ordering should give this
function a high priority to be eliminated first because this will
probably take the fewest steps to eliminate this function.

\section{Applications}  \label{exmpl} 
One of the applications to which the above length reduction procedure
has been applied successfully is the computation of Killing 
vectors $K_i(x^p)$ and rank 2 Killing tensors $K_{ij}(x^p)$ of
given space times with a metric $g_{ij}$ in General Relativity. 
Killing vectors and Killing tensors provide
conservation laws of the form $K_iu^i = $ const.\ and
$K_{ij}u^iu^j = $ const.\ where $u^i$ 
is the 4-velocity of geodesic motion in a curved space-time.
If enough Killing vectors (i.e.\ symmetries of the space time) and
Killing tensors are found then the equations of free
(geodesic) motion of test particles in a curved space can be integrated
which is a major step towards the physical interpretation of a space
time metric. Killing vectors and their Lie algebras are also widely
used to classify space time metrics.

The determining equations for Killing vectors are 
\begin{equation}
K_{i;j} + K_{j;i} = 0, \;\;\; i,j=1\ldots 4    \label{Keq1}
\end{equation}
and for Killing tensors they are 
\begin{equation}
K_{ij;l} + K_{jl;i} + K_{li;j} = 0, \;\;\; i,j,l=1\ldots 4    \label{Keq2}
\end{equation}
where `;' is the covariant derivative. Conditions (\ref{Keq1}),(\ref{Keq2}) are
generated for a given space-time with metric tensor $g_{ij}$ by the
program {\sc Classym} described in \cite{GreWo}. Examples
are shown in table 5 (not selected from a larger set, but as they appeared
in applications). The systems of equations determine either Killing
vectors (KV) or Killing tensors (KT) for different space time metrics.
One system is extended by integrability conditions (IC) and one system
by contractions (CT) with the metric tensor.
The number of equations and terms before and after length reduction are shown.
Running times in a {\sc Reduce 3.6} session running under {\sc Linux}
on a 133 MHz Pentium PC are given in the right column.

\begin{center}
\begin{tabular}{|l||c|c||c|c||c|c|} \hline
equations &\multicolumn{2}{c||}{original}
          &\multicolumn{2}{c||}{shortened} & no of & time \\ \cline{2-5}
          & eqn & terms & eqn & terms & red.\ steps & in sec \\ \hline \hline
KV for Taub-NUT (p170 in \cite{HE})
+ IC   & 49 & 1929 & 44 & 932 & 192 & 120    \\ \hline
KT for the Kerr metric (p161 in \cite{HE})
                       & 20 & 1154 & 20 & 750 &  36 &  87    \\ \hline
KT for the Kimura metric \cite{Kim}
+ CT & 48 & 464  & 21 &  74 &  93 &  19.5  \\ \hline
KV for the Barnes metric \cite{Bar}
                       & 10 &  66  & 10 &  39 &  14 &   2.3  \\ \hline
KV for pp waves (p178 in \cite{HE})
                       & 10 &  58  & 10 &  37 &  13 &   1.7  \\ \hline
\end{tabular} 
\end{center}

Table 5: The number of length reduction steps and the computation time for the
length reduction of different Killing vector (KV) and Killing
tensor (KT) determining systems of equations.\vspace{6pt}

The benefits of a length reduction
for a program like {\sc Crack} \cite{TW} are manifold.
Tables 6 and 7 show the number of terms and the occurring dependent
variables in each of the Killing tensor equations for the Kimura metric
before and after the reduction of length.

\begin{center}
\begin{tabular}{|c|c||c|c|} \hline
no of & dependent & no of & dependent \\
terms & variables & terms & variables \\ \hline
  2 & k01, k00           & 9 & k22, k11, k00, k13, k03, k33, k23 \\
  2 & k11                & 9 & k22, k11, k01, k00, k02, k03, k33 \\
  2 & k22, k12           & 9 & k22, k11, k01, k00, k02, k03, k33 \\
  3 & k13, k33, k23      & 9 & k22, k33, k23                     \\
  3 & k12, k00, k02      & 9 & k11, k01, k00                     \\
  3 & k00, k13, k03      & 9 & k11, k01, k00                     \\
  3 & k11, k01           &10 & k22, k12, k11, k00, k02, k33, k23 \\
  3 & k22, k01, k02      &11 & k22, k12, k13, k33, k23           \\
  3 & k12, k11           & 11& k22, k12, k13, k33, k23 \\
  3 & k11, k13           & 12& k22, k12, k11, k01, k00, k13, k33 \\
  4 & k01, k02, k03, k33 & 14& k22, k12, k11, k01, k13, k02, k03, k33 \\
  4 & k11, k01, k00      & 14& k22, k12, k11, k01, k13, k02, k03, k33 \\
  4 & k12, k01, k02      & 18& k12, k11, k01, k00, k13, k02, k03 \\
  4 & k01, k13, k03      & 18& k12, k11, k01, k00, k13, k02, k03 \\
  4 & k02, k03, k23      & 18& k12, k11, k01, k00, k13, k02, k03 \\
  4 & k22, k12, k11      & 18& k12, k11, k01, k00, k13, k02, k03 \\
  4 & k22, k13, k23      & 19& k22, k12, k11, k00, k13, k02, k03, k33, k23 \\
  5 & k22, k12, k33, k23 & 19& k22, k12, k11, k00, k13, k02, k03, k33, k23 \\
  5 & k12, k11, k13, k33 & 21& k22, k12, k11, k01, k00, k13, k02, k03, k33 \\
  5 & k12, k13, k23      & 21& k22, k12, k11, k01, k00, k13, k02, k03, k33 \\
  8 & k11, k01, k00      & 21& k22, k12, k13, k02, k03, k33, k23 \\
  8 & k12, k11, k00, k13, k02, k03 & 21& k22, k12, k13, k02, k03, k33, k23 \\
  8 & k12, k11, k00, k13, k02, k03 & 21& k22, k12, k13, k02, k03, k33, k23 \\
  8 & k11, k01, k00      & 21& k22, k12, k13, k02, k03, k33, k23 \\ \hline
\end{tabular} \vspace{6pt}\\
Table 6: The original set of conditions for Killing tensors in the
         Kimura metric.
\end{center}

\begin{center}
\begin{tabular}{|c|c||c|c|} \hline
no of & dependent & no of & dependent \\
terms & variables & terms & variables \\ \hline
  2 & k01, k00           & 4 & k22, k13, k33, k23 \\
  2 & k11                & 4 & k11, k01, k00      \\
  2 & k22, k12           & 4 & k01, k02, k03, k33 \\
  3 & k13, k33, k23      & 4 & k12, k01, k02      \\
  3 & k12, k00, k02      & 4 & k01, k13, k03      \\
  3 & k00, k13, k03      & 4 & k02, k03, k23      \\
  3 & k11, k01           & 4 & k22, k12, k11      \\
  3 & k22, k01, k02      & 5 & k22, k33, k23      \\
  3 & k12, k11           & 5 & k12, k11, k13, k33 \\
  3 & k11, k13           & 5 & k12, k13, k23      \\
  4 & k22, k33, k23      &   &                    \\ \hline
\end{tabular} \vspace{6pt}\\
Table 7: The length reduced conditions for Killing tensors in the
         Kimura metric.
\end{center}

The equations of table 7 (after dropping a single linear dependent
equation with 4 terms) are explicitly given in the appendix. It is
indicated there how the sparse occurrence of different derivatives
leads to integrable ODEs and exact differential equations.\vspace{6pt}

{\bf Length reduction as an adjunct to elimination algorithms}
\vspace{3pt}\\ 
The main purpose of the term reduction method is to shorten and
simplify systems of equations without imposing any risk of an
intermediate length increase. This risk is present when using standard
Gaussian elimination or standard Gr\"{o}bner basis methods.
For that reason and the benefits discussed above
it would not matter if the term reduction method would
be comparatively time consuming. The following run time tests show
that term reduction can even be 
time saving if it is used in connections with computing 
differential Gr\"{o}bner bases or the radical of a differential ideal.

The program {\tt rif} \cite{Reid2},\cite{Reid3} 
has been applied by Allan Wittkopf to the Killing tensor conditions
for the Kimura metric (see table 3, the shortened form is given
in the appendix). Because we only provide the relative speed up
in tables 8,9 the following hard and software specifications are not of
much relevance. They are added only for completeness.
Computing times of the program {\tt rif} are measured 
on a PII 333 MHz Linux system with 384 MB RAM (although only about 3 MB
was needed), running in {\sc XMaple} release 5.0.

Differential Gr\"{o}bner Bases have been computed, once in the generic
case where the constant $b$ is treated like a dependent variable and a
second time where $b$ is regarded as an arbitrary constant, ignoring
$b$ pivots. 

In the tables 8 and 9 kimu1 is the form of the system with 607 terms
(see table 3), kimu2 is a pre-optimal form of the system with 128 terms
which resulted when a non-optimal pairing of equations was used and
finally kimu3 is the system with 74 terms (see appendix). 

\begin{center}
\begin{tabular}{|c|c|c|c|} \hline
equations & {\tt rif}, with    $b$ pivots & 
            {\tt rif}, without $b$ pivots  \\ \hline \hline
$t_{\,\mbox{\small kimu2 }} /
 t_{\,\mbox{\small kimu1 }}$ 
                   & 96.8 \% & 93.2 \% \\
$t_{\,\mbox{\small kimu3 }} /
 t_{\,\mbox{\small kimu1 }}$    
                   & 86.6 \% & 72.3 \% \\     \hline
\end{tabular} \vspace{6pt}\\
Table 8: Relative speed up of computing times (cpu time) 
         for the program {\tt rif} for different
         versions of the Killing tensor conditions of the Kimura metric.
\end{center}

The {\sc Maple} program {\tt diffalg} has been applied by Evelyne Hubert
(see \cite{BLOP1, BLOP2, Hub, Hub2} and the URL
\verb+http://daisy.uwaterloo.ca/~ehubert/Diffalg+). {\tt
diffalg} was run on a PC with dual Intel 400Mhz CPU's, 
512 MB RAM, 128 MB swap, Asus P2L7-DS motherboard, 4.3 GB Seagate hard drive and
Intel EtherExpress Pro 10/100 network card under LINUX.

\begin{center}
\begin{tabular}{|c|c|} \hline
equations & {\tt diffalg} \\ \hline \hline
$t_{\,\mbox{\small kimu2 }} /
 t_{\,\mbox{\small kimu1 }}$ 
                   & 83.7 \% \\
$t_{\,\mbox{\small kimu3 }} /
 t_{\,\mbox{\small kimu1 }}$    
                   & 40.1 \% \\     \hline
\end{tabular} \vspace{6pt}\\
Table 9: Relative speed up of computing times (cpu time) 
         for the program {\tt diffalg} for different
         versions of the Killing tensor conditions of the Kimura metric.
\end{center}

A potential gain of a reduction of the number of terms is
the possibility to choose a more appropriate total ordering in a
Gr\"{o}bner Basis computations when a system is already partially
decoupled. For example, dependent variables which turn up in fewer
equations could
be given a higher lexicographical priority than other dependent
variables. So far no advantage has been taken of this opportunity
in the package {\sc Crack}.

\section{Possible extensions} \label{PosEx}
The manipulations studied in this paper to achieve a reduction of length 
seem very special and one might seek a more comprehensive theory, for
example, covering all possible linear combinations of any number of
equations multiplied not only with monomials but with any number of
terms. This problem is harder than it looks. It would include 
being able to decide whether any given linear algebraic system could be
solved without intermediate memory increase. Already the question of
combining 3 equations to kill at least 5 terms (4 terms could be
killed by combining 2 equations twice) has a much larger search
space than combining 2 equations. Although not proven, the author
expects that in generic practical applications the frequency of
reducing the length of any one of 3 equations which can not be reached
by combining repeatedly 2 equations is low (which, of course, may be
different in specific applications). To give an example, the
system $0=a+b-c-d+g,\;\; 0=c+d-e-f+h,\;\; 0=e+f-a-b+k$ gives a shortened
equation only if all three equations are combined (added).

It seems to be a general phenomenon that the computational complexity
increases drastically when any method of investigating systems of
equations is generalized whereas the success rate of the generalized
method increases only marginally. 

To give an example, in determining Lie-symmetries and conservation
laws of generic partial differential equations (not the rare fully
integrable PDEs) the success rate in finding symmetries when
going from point symmetries to higher order symmetries or the success
rate in finding conservation laws when going
from zeroth order integrating factors to higher order
conservation laws increases only marginally. On the other hand the
complexity in solving the related over determined system of conditions
grows drastically when the order of the ansatz is increased. The
growth in complexity comes from the task to determine an increased
number of constants or functions or in determining functions of more
independent variables. The decreasing success rate results from a
relatively more over determined problem: having to satisfy not only
the more restrictive conditions of a more general ansatz but also the
conditions that the solution may not be decomposable into simpler
cases. 


\section{Summary} 
Although conceptually simple, the method explained in this paper
proved to be very useful. It is fast and it has zero risk of increasing
the size of the system of equations during the computation or as a
result of it. This is an important feature for very memory intensive
computations. In such a case, computation times are of less concern.
Still, it was possible to show that even for other well known programs
the time saving using term reduction could be higher than the additional cost.

As discussed in the subsection `Other benefits', size reduced
equations on average are more likely to be an  
ODE or to be a total derivative and therefore easily integrable. 
They are also more likely to be algebraically factorizable.
The fact that the system becomes partially decoupled
opens the possibility to select a total ordering according to which
the system is already close to a characteristic form. This has the
potential to reduce subsequent computations to obtain a
characteristic system to a high extend. 

\section{Acknowledgement}
The author thanks Allan Wittkopf for many discussions. He, together
with Evelyne Hubert are thanked for running the tests
mentioned in the text with their systems. The Symbolic Computation Group at
the University of Waterloo is thanked for its hospitality during the authors
sabbatical in the fall of 1998 when the manuscript was prepared.

\section{Appendix: Length reduced conditions for Killing tensors of the
Kimura metric}
The size reduced determining conditions for Killing tensors in the
Kimura metric read:
\begin{eqnarray}
0&=&k_{22,h} + 2b^2r^3k_{12} \label{e1}  \\
0&=&rk_{11,r} + 2k_{11}  \label{e2}  \\
0&=&k_{00,t} - 2br^3k_{01}  \label{e3}  \\
0&=&2\cos(h)\sin(h)k_{23} + k_{33,p} + 2\sin(h)^2b^2r^3k_{13}  \label{e4}  \\
0&=&rk_{11,p} + 2rk_{13,r} - 2k_{13}  \label{e5}  \\
0&=&rk_{11,h} + 2rk_{12,r} - 2k_{12}  \label{e6}  \\
0&=&2k_{02,h} + k_{22,t} + 2b^2r^3k_{01}  \label{e7}  \\
0&=&2rk_{01,r} + rk_{11,t} - 2k_{01}  \label{e8}  \\
0&=&k_{00,p} + 2k_{03,t} - 2br^3k_{13}  \label{e9}  \\
0&=&k_{00,h} + 2k_{02,t} - 2br^3k_{12} \label{e10}   \\
0&=&6\cos(h)\sin(h)k_{23} - \sin(h)^2k_{22,p} - 2\sin(h)^2k_{23,h} +
    k_{33,p}  \label{e11}  \\
0&=&2\cos(h)\sin(h)k_{02} + 2k_{03,p} + k_{33,t} +
    2\sin(h)^2b^2r^3k_{01}  \label{e12}  \\
0&=&2rk_{12,h} + rk_{22,r} + 2b^2r^4k_{11} - 4k_{22}  \label{e13}  \\
0&=&2\cos(h)k_{03} - \sin(h)k_{02,p} - \sin(h)k_{03,h} -
    \sin(h)k_{23,t} \label{e14}  \\
0&=&rk_{01,p} + rk_{03,r} + rk_{13,t} - 4k_{03}  \label{e15}  \\
0&=&rk_{01,h} + rk_{02,r} + rk_{12,t} - 4k_{02}  \label{e16}  \\
0&=&rk_{00,r} + 2rk_{01,t} + br^5k_{11,r} - 4k_{00}  \label{e17}  \\
0&=&2\cos(h)\sin(h)^2k_{22} - 4\cos(h)k_{33} - \sin(h)^3k_{22,h} +
    2\sin(h)k_{23,p} \nonumber \\
 & &+ \sin(h)k_{33,h}  \label{e18}  \\
0&=&2\cos(h)\sin(h)rk_{12} + 2rk_{13,p} + rk_{33,r} + 2\sin(h)^2b^2r^4k_{11}
    - 4k_{33}  \label{e19}  \\
0&=&2\cos(h)rk_{13} - \sin(h)rk_{12,p} - \sin(h)rk_{13,h} - \sin(h)rk_{23,r} +
    4\sin(h)k_{23} \label{e20} 
\end{eqnarray}
One effect of reducing the number of terms is that fewer
different derivatives of different dependent variables $k_{ij}$ occur
in each equation. As a consequence the chance for the equations to be
exact or to be simple ODEs increases. The package {\sc Crack} is able to 
integrate 10 of the above equations and as a consequence to express
10 of the unknown functions $k_{ij}$ in terms of new unknown
functions of integration of only 3 variables: integration 
 of (\ref{e2}) to solve for $k_{11}$, of (\ref{e5}) to solve for $k_{13}$, 
 of (\ref{e6}) to solve for $k_{12}$, of (\ref{e1}) to solve for $k_{22}$,
 of (\ref{e8}) to solve for $k_{01}$, of (\ref{e3}) to solve for $k_{00}$,
 of (\ref{e9}) to solve for $k_{03}$, of (\ref{e7}) to solve for $k_{02}$, 
 of (\ref{e19}) to solve for $k_{33}$ and of (\ref{e20}) to solve for $k_{23}$.
These integrations and substitutions are a first step in
a longer calculation which finally gives the following general
solution. The 13 free constants $c_1\ldots c_{13}$ stand for 13
Killing tensors, i.e.\ 13 conserved first integrals of geodesic motion in the
curved space described by the Kimura metric. 
\begin{eqnarray*}
k_{00}&=&\left(3b^2r^4t^2c_{1} + 48b^2r^4tc_{2} + 3br^2c_{1} - b
r^4c_{3} - 3br^4t^2c_{4} - r^2c_{4}\right)/(6b)  \\
k_{01}&=&\left(brtc_{1} + 8brc_{2} - rtc_{4}\right)/(2b)  \\
k_{02}&=&\cos(p)r^4c_{12} + \sin(p)r^4c_{11}  \\
k_{03}&=&\left( - \cos(h)^3\cos(p)r^4c_{11} + 
        \cos(h)^3\sin(p)r^4c_{12} - \cos(h)^2\sin(h)r^4c_{13} \right.\\
& & \left.+ \cos(h)\cos(p)r^4c_{11} - \cos(h)\sin(p)r^4c_{12} + 
\sin(h)r^4c_{13}\right)/\sin(h)  \\
k_{11}&=& - c_{4}/(3b^2r^2)  \\
k_{12}&=&0  \\
k_{13}&=&0  \\
k_{22}&=&\left(6\cos(p)^2r^4c_{6} + 6\cos(p)\sin(p)r^4c_{5} - 3b^2r^4c_{1}
t^2 - 48b^2r^4tc_{2} + 3br^4t^2c_{4} \right.\\
& & \left.- 6r^4c_{7} - 2r^2c_{4}\right)/6  \\
k_{23}&=&\left(2\cos(h)^2\cos(p)r^4c_{8} - 2\cos(h)^2\sin(p)r^4c_{9} + 2\cos(h)
\cos(p)^2\sin(h)r^4c_{5}\right. \\
& & - 2\cos(h)\cos(p)\sin(h)\sin(p)r^4c_{6} 
- \cos(h)\sin(h)r^4c_{5} - 2\cos(p)r^4c_{8} \\
& &\left. + 2\sin(p)r^4c_{9}\right)/2  \\
k_{33}&=&\left(12\cos(h)^5\cos(p)r^4c_{9} + 12\cos(h)^5\sin(p)r^4c_{8} + 6\cos(
h)^4\cos(p)^2\sin(h)r^4c_{6}\right. \\
& & + 6\cos(h)^4\cos(p)\sin(h)\sin(p)
r^4c_{5} - 6\cos(h)^4\sin(h)r^4c_{7} + 6\cos(h)^4\sin(h)r^4c_{10}\\
& & - 24\cos(h)^3\cos(p)r^4c_{9} - 24\cos(h)^3\sin(p)r^4c_{8} - 
6\cos(h)^2\cos(p)^2\sin(h)r^4c_{6} \\
& &- 6\cos(h)^2\cos(p)\sin(h)\sin(p)r^4c_{5} 
+ 3\cos(h)^2\sin(h)b^2r^4t^2c_{1} - 12\sin(h)r^4c_{7} \\
& &+ 48\cos(h)^2\sin(h)b^2r^4tc_{2} - 3\cos(h)^2\sin(h)br^4t^2c_{4} - 6
\cos(h)^2\sin(h)r^4c_{6} \\
& &+ 18\cos(h)^2\sin(h)r^4c_{7} - 12\cos(h)
^2\sin(h)r^4c_{10} + 2\cos(h)^2\sin(h)r^2c_{4} \\
& &+ 12\cos(h)\cos(p)r^4c_{9} + 12\cos(h)\sin(p)r^4c_{8} 
- 3\sin(h)b^2r^4t^2c_{1} + 6\sin(h)r^4c_{10}\\
& & \left.- 48\sin(h)b^2r^4tc_{2} + 3\sin(h)br^4t^2c_{4} + 6\sin(h)
r^4c_{6} - 2\sin(h)r^2c_{4}\right)/(6\sin(h))
\end{eqnarray*}
What makes this result and the Kimura metric interesting is that two
of these Killing tensors are non-trivial, i.e.\ they are not just
symmetrized products of the 4 Killing vectors of this metric 
(see, for example \cite{Kim}).

\end{document}